\documentclass[11pt]{article}

\usepackage{Sty/standard}

\begin{document}

\title{High frequency trading and asymptotics for small risk aversion in a Markov renewal model\thanks{We would like to thank the AE and both referees for their numerous  comments, which were helpful for  improving the first version of this paper.}}

\author{\Fodra\and\Pham}

\maketitle

\begin{abstract}
We study a an optimal high frequency trading problem within a  market microstructure model designed to be  a good compromise between accuracy and tractability. The stock price is driven  by a Markov Renewal Process (MRP), as described in \cite{FodPha13a}, while market orders arrive in the limit order book via a point process correlated with the stock price itself.  
In this framework, we can reproduce the adverse selection risk, appearing in two different forms: 
the usual one due to big market orders impacting the stock price and penalizing the agent, and the weak one due to small market orders and reducing the probability of a profitable execution. 
We solve the market making problem by  stochastic control techniques in this semi-Markov model. 
In the no risk-aversion case, we  provide explicit formula for the optimal controls  and characterize the value function as a simple linear PDE.  In the general case, we derive the optimal controls and the value function in terms of the previous result, and illustrate how the risk aversion influences the trader strategy and her expected gain. Finally, by using a perturbation method, approximate optimal controls for small risk aversions are explicitly computed in terms of two simple PDE's, reducing drastically the computational cost and enlightening  the financial interpretation of  the results. 
\end{abstract}

\vspace{5mm}

\noindent {\bf Keywords: } High frequency trading, Markov renewal process,  Marked Cox process, adverse selection, 
integro-differential equation,  perturbation method. 
\vspace{5mm}

\newpage

\section{Introduction}

The existing literature on high-frequency trading  can be roughly divided into two main streams:  (i)  {\it Models of intra-day asset price}: this branch is devoted to the description of a tick by tick asset price in a limit order book, adopting 
two different philosophies. The \important{latent process} approach starts from a macroscopic unobserved process (typically a diffusion), which is contaminated by a noise reproducing the market  microstructure:  discreteness of prices valued in a tick grid, irregular spacing of price jump times (known as volatility clustering), and
mean-reversion of price variations. Some papers in this direction are \cite{GloJac01}, \cite{AitMykZha05}, and  \cite{RosRob11}.  
The \important{micro-macro} approach instead models directly the observed stock price by means of point processes, see  e.g. \cite{BauHau09},  \cite{ConDeL10}, \cite{AbeJed11} or \cite{BacHofMuz13}. 
 These papers consider sophisticated models, and are mainly  intended to reproduce microstructure stylized facts, as signature plot, Epps effect, volatility clustering and short mean-reversion. Often the main purpose is volatility 
 estimation, while trading applications are not studied: the complexity of these models leads to high dimensional equations in control problems, difficult to treat both analytically and numerically.  
 (ii) {\it High frequency trading problems}:  another important li\-terature stream focuses on trading problems in the limit order book: stock liquidation and execution problems (\cite{AlmCri00},  \cite{AlfSch10}, \cite{BayLud11}, \cite{guozer13}, etc ...),  
 or market making problems (\cite{AveSto08}, \cite{CarJaiRic11}, \cite{GuiPha13}, \cite{GueLeHTap12}, \cite{FodLab12}, \cite{CarWeb13}, etc ...). These papers use stochastic control methods to determine optimal trading strategies, and they are mostly based on  classical models for asset price, typically arithmetic or geometric Brownian motion, while the market order flow is usually driven by a Poisson process independent of the continuous price process.

 The goal of this paper is to make a bridge between these two streams of literature, by constructing a simple model for the asset price in a limit order book, intended to be both {\it realistic}, capturing the main stylized facts of microstructure, easy to estimate and simulate, and  {\it tractable}, (simple to analyze and implement) in order to lead to explicit formula in \hft\, applications with a nice financial interpretation. 
Since our point of view is the market maker's one, that we allow to interact with the market only by limit orders, \important{we will speak of limit orders only for those posted by the agent, while trades, or equivalently called market orders, refer to non-agent market orders arriving in the market}, and potentially matching the agent limit orders.


 We shall rely on our previous work \cite{FodPha13a}, where we show how  Markov Renewal processes provide  an extremely flexible and pertinent tool
to  model the stock price at high frequency, as well as easy to estimate, simulate and understand.  \important{We assume that the bid-ask spread is constantly one tick and the the stock price jumps of one tick}, which is consistent with liquid assets with large tick, see \cite{DayRos13}.  By modeling the stock price through a pure jump process (and not a continuous one as e.g. a Brownian motion), 
we are able to introduce probabilistic and mechanical dependences between the price evolution and the trades arrival.  We can easily introduce correlation between the next price jump and proportion of ask/bid trades before the next jump, as well as the jump risk.  In the context of  option pricing, jumps represent a source of market incompleteness, leading to unhedgeable claims.
Similarly, jumps of the stock price in the electronic market are a real source of risk for the market maker. More precisely, the agent faces two kinds of risk:  (i)  {\it Market risk:} when the price suddenly jumps, the whole agent inventory is re-evaluated, changing the portfolio value in no time (i.e. a finite amount of risk in no time, whereas the Brownian motion has quadratic variation proportional to the interval length). 
(ii)  {\it Adverse selection risk:} in our model, we assume that an upwards (resp. downwards) jump at time $t$ corresponds to a big market order clearing
the liquidity on the best ask (resp. bid) price level.  If the agent has posted a \important{small} limit order, say on the bid side, the latter has to be executed, since  the goal of the big market order on the bid side is in principle to clear all the available liquidity rather than consuming a fixed amount of it. In this sense, the agent does not affect the market dynamics.  In this scenario, the agent is systematically penalized, since she sells liquidity at $t^-$ for less than its value at $t$. This risk, known as \important{adverse selection} in the market microstructure literature, see e.g. \cite{ohara97}, will be incorporated, measured and hedged, in our market model.


 We introduce a suitable modeling of the market order flow, taking into account the real dependence with the stock price dynamics. 
The existing literature has provided several models for the arrival of trades in the limit order book. The seminal work  \cite{EngRus98} describes these events as a time series with an autoregressive behaviour, in order to model intensity spikes in the trading activity. Other authors (e.g. \cite{CarJai13}), exploit renewal processes for modeling trades and describe the price formation in terms of this flow. One of the richest example is provided by \cite{BacMuz13}, where a four-dimensional Hawkes process drives both trades and price, taking into account the mutual interaction of all its components: unfortunately, this elegant approach leads to a high-dimensional system, and consequently to computational issues. 
We mention also \cite{CarJaiRic11} in which a generalised multi-variate Hawkes process for activity arrivals and midprice movements is used for solving a market making problem, as well as \cite{Lar07} where the author performs a thorough statistical analysis by means of a ten-variate Hawkes process. 
\important{Our model is price rather than trade centered}, since we assume that trades (no matter their side) are counted by a Cox process subordinated to the stock price. Trades arrive more frequently after a price jump, while their arrival rate decreases as the price stabilizes. In this sense, events have  a much richer dynamic than a Poisson process, and they are not independent from the stock price, as often assumed. For this improvement, we pay no computational cost: we will see that the state variables describing the stock price are all what we need.  We do not include  self-exciting components as in Hawkes process  in order to keep the model dimension to the minimum. 


  By  adding marks (determining the trade exchange side) to the Cox process counting the trade events, we are able to reproduce in addition to the adverse selection already mentioned in this introduction, another form 
of risk limiting the agent profit, and called  {\it weak  adverse selection}. 
It comes from the small trades flow, made of those trades that are unable to move the market stock price. For this flow, a limit order is more likely to be matched on the less profitable side: if the price is likely to jump downwards (resp. upwards), few trades would arrive at the best ask (resp. bid) price, limiting the chances of building a short (resp. long) position (that would be profitable w.r.t. to the market direction). Thanks to this feature,  the extra gain (w.r.t. to a market order) due to a limit order execution is compensated by an unfavorable execution probability.

 In this context, we study the market making problem of an agent submitting optimally 
limit orders (market orders, as in \cite{GuiPha13}, are not taken into account to improve the problem tractability) at best bid and best ask prices. The agent is not the unique market maker, but only one of the many participants of the exchange, and she has no constraint in terms of liquidity providing. 
Several authors in previous literature, as e.g. in the seminal paper \cite{AveSto08},  consider limit orders, which are posted at a mid-price distance  
which may be nonpositive (some exception is the paper \cite{CarJaiRic11} which imposes  a nonnegative constraint on the distance).  This  induces some problems with high frequency applications, in particular for large tick assets (see \cite{DayRos13}) or pro-rata limit order book, where most of the liquidity is concentrated in few levels, making real-life rounding of the optimal quotes to the tick grid a non trivial issue.  
By replacing real controls with binary ones (limit order placed or not at the best price) we artificially add a policy constraint, but in such a specific context,  
the model fits better the reality: estimation is easier, policies are not subject to rounding as in the real case, no problem of negative posting distance exists, and the market spread is never improved;  in other words \important{theoretical policies need  no translation to real-trading ones}.  In this framework, we are able to derive explicit formula for both the value function and the (binary) optimal controls for an agent without risk aversion. Next, by  a perturbation technique, we solve the market making problem for small risk aversions. This  allows us  to reduce drastically the problem dimension, and greatly improves  the financial interpretability of the optimal strategy. In particular, we clearly understand the potential sources of both gain and risk in the model, and how the introduction of the risk aversion deforms them.


The structure of the paper is organized as follows.  We briefly review in Section \ref{Stock price in the limit order book} the asset price model introduced  in \cite{FodPha13a}, and derive some useful results concerning the conditional  mean of the stock price, i.e. a trend indicator.  In Section \ref{section::Market order flow modeling}, we describe the market order flow modeling small market trades, i.e. those trades leaving the stock price unchanged. On the contrary, big market orders, i.e. those affecting immediately the stock price value, are assumed to be incorporated into the price dynamics. For the small trades,  we introduce a marked Cox process subordinated to the stock price dynamics in order to reproduce the weak  adverse selection risk and the intensity correlation between the two processes. Section \ref{The market making problem} includes the formulation of the market making problem, and describes the agent process (wealth, inventory) dynamics, 
while Section \ref{section::The value function and the optimal controls: a perturbation method approach} is devoted to the resolution of the market making problem,  by using a perturbation method for small risk aversion.   Some numerical tests illustrate our results.  We conclude in Section 6, and collect some useful results in the Appendix.

\subsubsection*{Notation}

From now on, in order to keep notations compact, we will use, for any function ${f=f(x)}$, $x\in\R^n$, the following notation:
\begin{eqnarray*}
\Delta_x f(x') := f(x') - f(x) 
\end{eqnarray*}
and we shall omit the subscript $x$ in $\Delta$ $=$ $\Delta_x$ when there is no ambiguity from the context.

\newsection{Stock price in the limit order book}  \label{Stock price in the limit order book}

We consider  a model for the mid-price of a stock (the arithmetic  mean between the best-bid  and best-ask price) in a limit order book (LOB) with a constant bid-ask spread ${2\delta>0}$. For simplicity, we assume that the stock price jumps only of one tick. 
The mid-price $(P_t)_{t\geq 0}$ is then defined by the c\`ad-l\`ag piecewise constant process
\beq \label{midprice}
P_t &:=& P_0 + 2 \delta \sum_{n=1}^{N_t} J_n ,
\enq
where  $P_0$ is the opening mid-price valued in $2\delta\Z$, 
$N_t$ (representing the \important{tick times}) is the point process associated to the price jump times $(T_n)_n$,  i.e. $N_t$ $:=$ $\inf\big\{n: \sum_{k=1}^n T_k \leq t \big\}$, 
and  $J_n$ (called  \important{price direction}) is the marks sequence valued in $\pmset$ indi\-cating whether the price has jumped  upwards ($J_n=+1$) or downwards ($J_n=-1$) at time $T_n$.

\subsection{Markov renewal model}

We  use a  Markov renewal approach  as in \cite{FodPha13a} for modeling the marked point process $(T_n,J_n)_n$, and briefly review the main features.   The \important{price direction} is 
driven by the Markov chain 
\begin{eqnarray}
J_n &=& J_{n-1} B_n \; =\;   J_0 \prod_{i=1}^{k} B_i,  \for n\geq 1, \labeleq{Bn}
\end{eqnarray}
where $B\equiv(B_n)$ is an i.i.d. sequence, independent of $(J_n)$, and distributed according to a Bernoulli law on $\pmset$ with parameter  
$(1+\alpha)/2$, $\alpha\in[-1,1)$. In other words, the probability of two consecutive jumps in the same (resp. opposite)  direction is given by
\begin{eqnarray}
\condprob{P}{}{J_nJ_{n-1}=\pm 1}&=&\frac{1\pm\alpha}{2}\for  n\geq 1, \nn
\end{eqnarray}
where $\alpha$ represents the correlation between two consecutive price directions $J_{n-1}$ and $J_n$ under the stationary measure $\pi=(1/2,1/2)$ associated to the Markov chain 
$(J_n)$.  For  $\alpha=0$,  the jumps of the stock price are independent, for $\alpha>0$,  the price is short-term trended, while for $\alpha<0$,  the stock price exhibits 
a short-term mean-reversion, which is a  well-known stylized fact about high-frequency data, usually called microstructure noise.

In a second step,  we  model the  counting process $(N_t)$. Denoting by 
\begin{eqnarray*}
S_n &:=&T_n-T_{n-1}\geq 0\for  n\geq 1, \labeleq{elapsed_time}
\end{eqnarray*}
the inter-arrival times of the price jump times, we assume that, conditioned  on $J_n J_{n-1}$,
$(S_n)$ is an independent sequence of random variable with  distribution
\beqs
F_\pm(s) &:=& \condprob{P}{}{S_{n+1} \leq s \condto J_n J_{n-1} = \pm 1} \for  n\geq 1, s\geq 0,
\enqs
and density $f_\pm(s)$ with no masses.  We can easily check that $(S_n)$ is also unconditionally i.i.d., which implies that $(N_t)$ is the renewal process with inter-arrival times distribution given by
\beq \label{defF} 
F &:=& \frac{1+\alpha}{2} F_+ + \frac{1-\alpha}{2} F_-\stop
\enq
We define the pure jump process valued in $\pmset$
\begin{eqnarray*}
I_t &:=& J_{N_t},  \;\;\;\;\;  t\geq 0, \labeleq{I}
\end{eqnarray*}
which gives at time $t$ the direction of the last jump of the stock price. $(I_t)$ is a semi-Markov process in the sense that the pair $(I_t,S_t)$ is a Markov process, where 
\begin{eqnarray}
S_t &:=& t- \sup\{T_n:  T_n \leq t \} \; \geq 0 \;  \for t\geq 0, \labeleq{elapsed_timet}
\end{eqnarray}
is the elapsed time since the last  price jump.   Finally, we set to
\begin{eqnarray*}
\jumpintensityand\pm & := & \lim_{\Delta s\rightarrow 0^+} \frac{1}{\Delta s} \condprob{P}{}{s \leq S_{n+1} \leq s + \Delta s, J_{n+1} = \pm J_n \; \condto \;   S_n \geq s, J_n}  \\
&=& \frac{1\pm\alpha}{2}\,\frac{f_\pm(s)}{1-F(s)}\for s \geq 0,  \labeleq{hpm}
\end{eqnarray*}
the intensity function of price jump in the same (resp. opposite) direction, assumed to be a bounded continuous function,   
and  define $\drift$ (resp. $\vol$) as the measure of the conditional trend (resp. agitation state) of the stock price: 
\begin{eqnarray*}
\drift \; := \;  \jumpintensityand+ - \jumpintensityand-, \labeleq{mu} & &  
\vol \; := \;  \jumpintensityand+ + \jumpintensityand-  \;  \geq \;  0 \labeleq{sigma} \stop
\end{eqnarray*}
We recall that the elapsed time process $(S_t)$ is an homogeneous Markov jump process with stochastic intensity $\sigma^2(S_t)$ (the same as the renewal process $N_t$, since they jump at the same time) and infinitesimal generator:  $\varphi(s)$ $\mapsto$ $\Ds{\varphi} + \vol [\varphi(0)- \varphi(s)]$.

 \subsubsection*{Data sample}
We refer  to \cite{FodPha13a} for the statistical estimation of the trendiness parameter $\alpha$, the distribution $F_\pm$ of the renewal times, and the jump intensities $\jumpintensityand\pm$.  In the sequel, market data are taken from tick-by-tick observation of the 3-month future EUROSTOXX50, on February 2011, from 09:00:000 to 17:00:00.000 (CET).   Furthermore, since $(N_t)$ is a renewal process associated to the renewal distribution $F$ given in \reff{defF}, we will often refer, in order to obtain a natural rescale for $s$, to
\begin{eqnarray*}
\hat s &:=& F^{-1}(s) \for s\geq 0\labeleq{renewal_quantile}
\end{eqnarray*}
as the renewal quantile associated to the elapsed time $s$. In order to increase the interpretability of the results, all the charts in this paper rescale the state variable $s$ according to this transformation.
Figure \ref{fig::hazard} plots the form of $\jumpintensityand\pm$ as a function of the renewal quantile: the reverting intensity $\jumpintensityand-$ is dominant w.r.t. to the trending intensity $\jumpintensityand+$ for small renewals, while  this discrepancy tends to disappear for higher quantiles ($>0.50$),  ending with $\jumpintensityand+$ dominating $\jumpintensityand-$ on the right boundary.
This tells that when the price is stable, i.e. its quotation has been constant for a relatively long time, the micro-structural mean reversion disappears, and the price looks like a Bernoulli random walk.

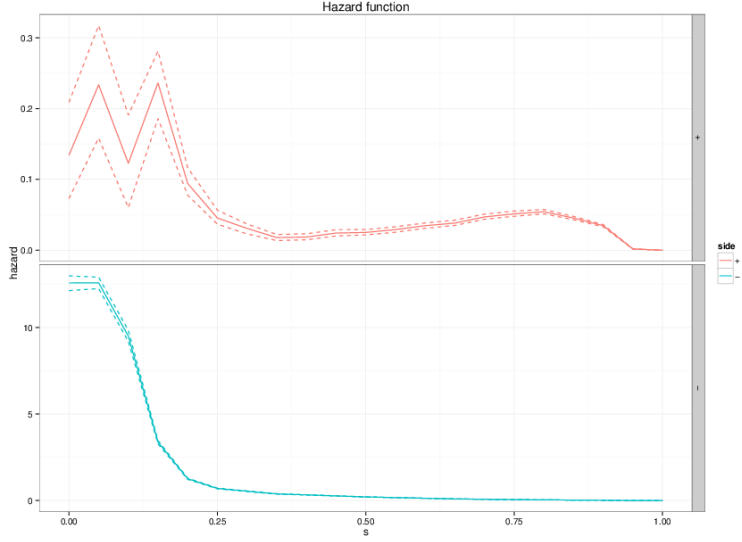
\newfigure{hazard}{Non parametric estimation of $\jumpintensityand\pm$ as a function of the renewal quantile ($\alpha=-0.75$), with bootstrap 95\% confidence intervals.\ep}

\subsection{The stock price conditional mean and the trend indicator} \label{sectheta} 

We recall that the price process $(P_t)$ is embedded into a Markov system with three obser\-vable state variables:  $(P_t,I_t,S_t)$ is a Markov process with infinitesimal generator: 
\beq \label{infiniP}
\varphi(p,i,s)  \;  \mapsto \; \Ds{\varphi} + \sum_{\sym\in\pm} \jumpintensityand\sym \Delta\varphi(p+2\sym\delta i,\sym i,0).  
\enq
In the sequel, for the applications of our model to market making problem, we shall extensively use properties of the mean value of the stock price at horizon, i.e. 
\beq \label{pi}
\pi(t,p,i,s) &:=&\condprob{E}{t,p,i,s}{P_T}.   
\enq
Here, $\mathbb{E}_{t,p,i,s}$ denotes the expectation operator under the initial conditions 
$$
(P_{t},I_{t},S_{t}) \; = \; (p,i,s), \;\;\;  (t,p,i,s)\in[0,T]\times 2\delta\Z \times \pmset \times \Rplus.
$$
We devote a separate part to the study of this function, which is fully developed in the Appendix \ref{appendix::thetaT}. Here we concentrate the main results concerning $\pi(t,p,i,s)$ in the following proposition, which will be a useful reference in the remaining part of the paper.

\begin{Proposition}\label{prop::theta}
The function $\pi$ in \reff{pi}  is given by
\begin{eqnarray}
\pi(t,p,i,s)  &=& p +  i \theta(t,s) , \label{ansatzmean_price}
\end{eqnarray}
where $\theta(t,s)$ is  the unique   bounded continuous  viscosity  solution to
\begin{equation} \labeleq{theta}
\begin{cases}
- {\cal M}\theta
-  2 \delta \drift \; := \;  - \Dt{\theta}  - \Ds{\theta}  - \drift \theta(t,0) +  \vol \theta(t,s) -  2 \delta \drift \; = \;  0  ,\\ 
\theta(T,) \; = \; 0,
\end{cases} 
\end{equation}
for $(t,s)\in[0,T]\times\Rplus$.  Furthermore,  fixed $T\geq 0$,  $\theta(t,s):=\theta^T(t,s)$ satifies
\begin{eqnarray}\labeleq{theta_infty}
\lim_{T\rightarrow\infty} \theta^T(t,s) \; = \;  2\delta \frac{\tilde\alpha(s)}{1-\alpha} \;  =:  \; \theta^\infty(s) \for \forall (t,s) \in\R_+^2,
\end{eqnarray}
where 
\begin{eqnarray*}
\labeleq{tilde_alpha}
\tilde\alpha(s) &:= & \condprob{P}{}{J_1\condto  I_{0} =+1,S_0 = s} \; = \;  \sum_{\sym\in\pm} \sym\left(\frac{1+\sym\alpha}{2}\right)\left(\frac{1-F_{\sym}(s)}{1- F(s)}\right), \label{tilde_alpha}
\end{eqnarray*}
with  $\tilde\alpha(0)=\alpha$ and  $\tilde\alpha(s)\equiv\alpha$ if marks and tick times are independent.
\end{Proposition}
\begin{proof}
See Propositions  \ref{lemma::theta},   \ref{cor::tilde_alpha} and   \ref{prop::phi_infty}.
\end{proof}

\begin{Remark}
{\rm 
From the definition of the function $\pi$, and relation \reff{ansatzmean_price}, we see that $\theta$ $=$ $\thetaT$ admits  the probabilistic representation:
\begin{eqnarray} \label{thetaproba}
\thetaT(t,s) &=& \condprob{E}{}{P_T\condto  P_t = 0, I_t = +1,S_t = s},
\end{eqnarray}
and can be interpreted as a martingale deviation of the stock price ($\theta$ $\equiv$ $0$ means that price is martingale). \ep
}
\end{Remark}



\newsection{Market order flow modeling and adverse selection} \label{section::Market order flow modeling}

In this section, we  model the market order flow, that is the counterpart trade of the limit orders in a LOB. We distinguish two types of market orders: 
\begin{enumerate}
\item \important{big market orders} that move the bid or ask price, and so induce jumps in  the mid-price  according to the mechanism described in the previous section;
\item \important{small market orders} that do not move price. 
\end{enumerate}

Let us first describe the impact of big market orders. Suppose that a limit order is posted, say at the bid price. If a big market order arrives at the bid price, and consumes all the available liquidity, the price jumps downwards. The agent limit order is then executed at the new best ask price, i.e. it becomes a market buy order, hence is executed unfavorably. This phenomenon is known in the literature on market microstructure 
as the  \important{adverse selection}.  We discuss more in detail this feature in the next section, see Remark \ref{remstrongadverse}.

We now focus on  small market orders that we model by a marked point process $(\theta_k,Z_k)$.
\begin{enumerate}
\item {\it Timestamps:} the increasing sequence  $(\theta_k)$ represents the 
arrival  timestamps of (small) market orders.
\item {\it Marks:} the marks $(Z_k)\in\pmset$, represent the side of the exchange, with the convention that when $Z_k= -1 \;(\text{resp. } +1),$
the trade is exchanged at the best bid (resp. ask) price, i.e. market sell (resp. buy) order has arrived. 
\end{enumerate}
In this paper, we do not take into account the size of trades. 

\subsubsection*{The market order counting process.} On one hand, we assume that the counting process $(M_t)$ associated to $(\theta_k)$ is a Cox process with conditional intensity $\lambda(S_t)$, where  $\lambda$ is a bounded continuous function on $\R_+$.  This approach extends the simple case where $(M_t)$ is a Poisson process. For the estimation of $\lambda$, we have used a parametric approach based on the MLE algorithm for point processes 
(see e.g. \cite{andetal93}), as shown by Figure \ref{fig::lambda_quantile}, where $\lambda$ is described as a function of the renewal quantile.  When the price is very unstable, many trades arrive in the limit order book. On the contrary, when the price stabilizes, the trading activity is weaker, but present.

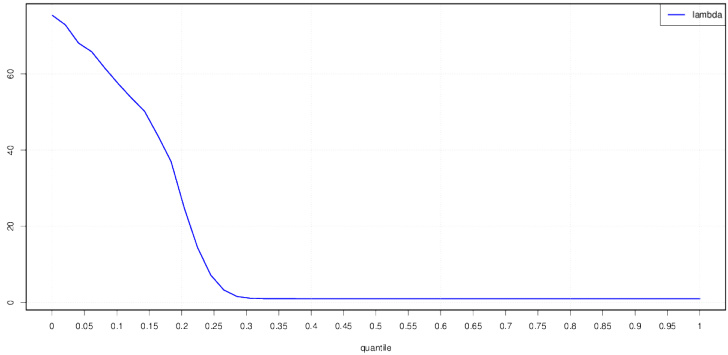
\newfigure{lambda_quantile}{estimation of $\lambda(s)=\lambda_0 + a\exp(-ks)$. The conditional rate of order arrival decays exponentially with the time passing with no event occurring, but never completely disappears, as shown by $\lambda_0>0$, representing the base and minimal intensity of the process.\ep}


\subsubsection*{The market order exchange side.} On the other hand, we assume that  market order trade and stock price in the LOB are correlated via the relation: 
\beq \label{modZ}
Z_k &:=& \Gamma_k \; I_{\theta_k-},
\enq
where $\Gamma\equiv(\Gamma_k)$ is an i.i.d. sequence, independent of all other processes, and distributed according to a Bernoulli law on $\pmset$ with  parameter $(1+\rho)/2$, with $\rho$ $\in$ $\left(-1,1\right)$. 
Since
\beqs
\rho &=& \text{Corr}(Z_k,I_{\theta_k^-}),
\enqs
\begin{enumerate}
\item for $\rho = 0$, the trade sides do not depend on the stock price, and market order flow arrive independently at best bid and best ask. This is the usual assumption made in the literature, see e.g. \cite{AveSto08};
\item for $\rho > 0$, most of the trade sides are \important{concordant} with the direction of the last jump of the stock price: market orders arrive more often in the 
\important{strong}  side ($+$) of the limit order book, i.e. the side in the same direction than the last jump - best ask (resp. bid)  when price jumped upwards (resp. downwards);
\item for $\rho<0$, most of the trade sides are \important{discordant} with the direction of the last jump of the stock price: market orders arrive more often in the 
\important{weak}  side ($-$) of the limit order book, i.e. the side in the opposite direction than the last jump - best bid (resp. ask)  when price jumped upwards (resp. downwards). 
\end{enumerate}
We define, consistently with the notation for  $\jumpintensityand\pm$, the concordant (+) and discordant (-) trade intensity as
\begin{eqnarray}
\tradeintensityand\pm &:=& \left(\frac{1\pm\rho}{2}\right) \lambda(s)\for s\geq 0 \labeleq{lambdapm}. 
\end{eqnarray}

\subsubsection*{The weak adverse selection risk.}

Using the the strong law of large numbers,
estimation on real data leads to a value of $\rho$ around $-50\%$. This means that about $3$  over $4$ trades arrive on the weak side of the limit order book.  Recall that stock price usually  
exhibits a short-term mean reversion, i.e. a negative correlation $\alpha$ of price increments, so that  $\alpha \rho > 0.$
The fact that $\alpha$ and $\rho$ are of the same sign is consistent with the execution dynamics. Suppose on the contrary that $\alpha$ and $\rho$ are of opposite sign, say 
$\alpha<0$ and $\rho>0$, and assume e.g. that the last price jumped downwards. Then $\rho>0$ means that most of the trade will occur at best bid, hence will execute limit buy orders in a bull market (since $\alpha<0$), allowing market makers on best bid to open a low risk profitable position. The quantity $\alpha\rho$ is a measure of the inefficiency of limit orders w.r.t. to trend capturing: the bigger it is, the lower is the probability of building a profitable position via a limit order. This quantity compensates the intrinsic advantage of a limit order execution (one spread w.r.t. to a market order), and provides a first explanation of why market making is not a trivial game. We call this phenomenon the \important{weak adverse selection}, in comparison with the usual  adverse selection  considered above  (see also Remark \ref{remstrongadverse}) and related to big market orders.  

\newsection{The market making problem} \label{The market making problem}

\subsubsection*{The agent strategy.}

The agent strategy consists in placing continuously  limit orders of constant small size $L\in\N\setminus(0),$ 
(where small is meant w.r.t. to the total liquidity provided by all the market makers) on both sides, and  at the best price available.   
The market making strategy is then described by a pair of predictable processes   $(\ell^+,\ell^-)$ valued in $\{0,1\}$.   
When $\ell^+_t=1$ (resp. $\ell^-_t=1$), a limit order of  size  $L$  is posted at time $t$ on the strong (resp. weak) side, while in the opposite case
no limit order is submitted or the limit order is cancelled.  We denote by $\Ac$ the set of market making controls {\bf $\bell$} $=$ $(\ell^+,\ell^-)$.

Every time  a small  market order arrives in the limit order book (according to the mechanism described in Section 3),  
if the agent  has placed an order on the corresponding side,  the latter is  executed according to a random variable, whose distribution may differ according to the book matching rules. We recall the two main frameworks we can deal with.
\begin{enumerate}
\item {\it Price time priority (PTP)}:  in a PTP order book, orders are matched according to their price (more priority to those closer to the mid price) and time (like in single file queue). 
\item {\it Pro-rata}: in a (pure) pro-rata order book, there is price priority (as in PTP), but not the time one. Multiple market makers are executed when a single trade arrives, according to a proportion  rewarding the limit orders of bigger size. 
\end{enumerate}

Notice that, if the price jumps upwards and the agent is in the condition to place her limit orders immediately after the jump, she will find a huge volume of concurrent limit order on the ask side (old orders), while very few on the bid side. The situation is symmetric when the price has jumped downwards.  We then consider the distributions
\beqs
\vartheta_\pm(dk,L) & \mbox{ on } & \{0,\ldots,L\}, 
\enqs
where $\vartheta_-(dk,L)$ (resp. $\vartheta_+(dk,L)$)  is  the distribution of  the executed quantity of limit order of size $L$ in the concordant 
(resp. discordant), i.e. strong (resp. weak)  side, of the limit order book (an estimation procedure for $\vartheta_\pm$ is provided in the Appendix \ref{appendix::estimation_vartheta}). We  denote by 
\beq \label{momentvartheta}
\vartheta_\pm^m(L)  &:=& \int  k^m \vartheta_\pm(dk,L)\for m=1,2,
\enq
the first and second  moment of these distributions.

\begin{Remark}[The adverse selection risk]\label{remstrongadverse}
Recall once again that  the sign $\pm$ indexing the distribution must not be interpreted as the ask/bid side of the limit order book but as the  strong/weak side of the limit order book taking into account the last price direction. 
Moreover, in order to be consistent with the small agent assumption (the price is exogenous and not impacted by the agent strategy), we assume that, if the price jumps across  the level of a limit order placed by the agent, the latter is automatically executed. This can be justified by saying that, since the agent is small, the price  jumps  independently of the presence of her limit order, since a  large market order whose goal was  (at least) to  clear the best level. Furthermore, since the spread is constantly one tick and orders are placed at the best price, when the price jumps,  limit orders converts automatically to market ones, which are immediately executed. Notice that this scenario is particularly adverse to the agent: she is  selling (resp. buying) a stock at ask price $P_{t-} + \delta$ (resp. bid price $P_{t^-}-\delta$), while the current mid-price is $P_t = P_{t-}+2\delta$ (resp. $P_t = P_{t-}-2\delta$ ), so in both cases she is losing $\delta$ for each lot! The sudden execution is both against the market and does not let her the time to close the spread, leaving an open position: she faces both adverse selection and inventory risk.   
This phenomenon, known as  \important{adverse selection}, is considered also in \cite{CarJaiRic11} for example, where the authors consider a Brownian stock price with stochastic mean reverting drift (called $\alpha$), which is impacted by incoming influential trades. In this paper, an influential trades on the ask (bid) side, makes the stochastic drift jump upwards (downwards), reproducing the intrinsic  connection between the trade flow and the dynamic of the stock price.  
\ep
\end{Remark}

\clearpage

\subsubsection*{The wealth and the inventory process.}

We assume that each transaction is subject to a fixed cost $\eps$ $\geq$ $0$.   Let us then denote  by $(X_t)$ and $(Y_t)$ the wealth and inventory describing the agent portfolio, valued respectively in $\R$ and $\Z$.

\begin{Lemma} \label{lemXY}
For a market making strategy $\bf \bell \in \Ac$,  the dynamics of the portfolio value processes $X$ and $Y$ are given by
\begin{eqnarray*}
dX_t 
&=& \sum_{\sym\in\pm} \int k \; \ell^\sym_{t-}\left(\sym P_{t-}I_{t-} + \delta-\eps\right)\;\left(R_\sym^{trd}\left(dt,dk,S_{t-}\right)+R_\sym^{jmp}\left(dt,dk,S_{t-}\right)\right), \\
dY_t 
&=& -\sum_{\sym\in\pm} \int k\, \sym\,\ell^\sym_{t-} I_{t-}\left(R_\sym^{trd}\left(dt,dk,S_{t-}\right)+R_\sym^{jmp}\left(dt,dk,S_{t-}\right)\right), 
\end{eqnarray*}
where $R_\pm^{\text{trd}}$ (resp. $R_\pm^{\text{jmp}}$) is  a  random measure  with intensity
$$ 
\lambda_\pm(S_{t^-})\;dt\otimes\vartheta_\pm(dk,L)\;\;\;
(resp. \; \jumpintensity_\pm(S_{t^-})\;dt\otimes  {\bf \delta}_{L}(dk)),
$$
and  ${\bf \delta}_{L}$ is the Dirac measure at $L$. 
\end{Lemma}

\begin{proof}
To fix the ideas, assume that, at time $t$, $i=I_{t-}=+1$, i.e. the last jump of the price has been upwards (in the case $i=-1$ the proof is symmetric). Then if $\ell^\pm_{t^-} = 1$, i.e. if the agent has placed a limit order on the strong (i.e. ask) or the weak (i.e. bid) side: 
\begin{enumerate}
    \item either she can be executed by an incoming trade of a random number of quantities $k$, where $k$ has distribution $\vartheta_\pm(dk,L)$, with 
     trade intensity  $\lambda_\pm(S_{t^-})$
	\item or she can be entirely executed, thus $k=L$, by a jump of the price, with rate $h_\pm(S_{t^-})$.
\end{enumerate}
In both cases, the agent inventory $Y$ jumps of $\mp k$, while her wealth $X$ jumps of $k$ times $\pm P_{t-}$ (asset sold or bought), plus the market making prime of $\delta$ (the half spread), minus the fixed cost $\eps$.
\end{proof}

\begin{Remark} \label{remQ}
{\rm It will be convenient in the sequel to introduce the process:  $Q_t$ $:=$ $I_t Y_t$, called strong inventory process. Since $I_t$ valued in $\{-1,1\}$ jumps only when there is a price jump on the weak side $\nu$ $=$ $-$, 
we then see from the dynamics of $Y$ that given $\bf \bell \in \Ac$, $Q$ evolves according to: 
\beqs
dQ_t &=& -\sum_{\sym\in\pm} \int k\, \sym\,\ell^\sym_{t-}  R_\sym^{trd}\left(dt,dk,S_{t-}\right) + 
\sum_{\sym\in\pm}  \int \big( (\nu-1)Q_{t^-} - k \; \ell^\sym_{t-} \big) R_\sym^{jmp}\left(dt,dk,S_{t-}\right). 
\enqs
\ep
}
\end{Remark}

\subsubsection*{The value function}
We now consider the following criterion for the market making optimization problem, as usually adopted in \cite{AveSto08}, \cite{CarJai13}, \cite{GuiPha13}: 
the agent is looking for the optimal admissible market making strategy, which maximises her expected portfolio value at terminal date $T$, evaluated at the mid price, while controlling her final inventory through a quadratic penalization term  $\eta Y_T^2$, with a risk aversion parameter $\eta$ $\geq$ $0$.

We  can now define in our MRP model  the value function associated to the market making problem: 
 \begin{eqnarray} \label{valuefunction}
v(t,p,i,s,x,y) &:=&  \max_{\bell \in \Ac} \condprob{E}{t,p,i,s,x,y}{X_T + Y_T P_T-\eta Y_T^2},
\end{eqnarray}
for $(t,p,i,s,x,y)$ $\in$ $[0,T]\times 2\delta\Z \times \pmset \times \Rplus \times\R\times\Z$, and
where  $\E_{t,p,i,s,x,y}$ denotes the expectation operator under the initial conditions $P_{t}=p$, $I_{t}=i$, $S_{t}=s$, $X_{t}=x$, $Y_{t}=y$.

The choice of a linear/quadratic utility function rather than an exponential one is purely due to computational ease: the utility function being polynomial in the state variable $Y$ allows us  to split PDE's according to their degree, reducing the problem dimension. The penalization term $-\eta Y^2_T$  limits the agent inventory risk: the agent does not want to close her trading day with a large position, that she would execute via a unique market order impacting the market. The quadratic form is consistent with a limit order book with constant liquidity shape, as explained in \cite{CarJai12}: the risk aversion parameter 
$\eta$ has to be considered as the subjective aversion of the agent to a final market order. This risk term was also recently demonstrated to stem from 
model ambiguity, see \cite{cardonjai13}.  Since trading horizons are usually short (minutes), this penalisation affects the whole strategy, and provides an excellent inventory control throughout the whole trajectory, as we will illustrate later. 

As an alternative, this penalisation can be replaced by  $-\eta\int_t^TY_u^2du$, without changing substantially the arguments we are going to adopt to solve the control problem. 
However, while penalization on final inventory leads to  semi-explicit calculations as we shall see in the next section, this is 
no more the case when considering penalization over the whole trading time interval.

\newsection{Value function and optimal controls: a perturbation approach} \label{section::The value function and the optimal controls: a perturbation method approach}


From the expression \reff{infiniP} of the infinitesimal generator of $(P,I,S)$, and the dynamics of the controlled process $(X,Y)$ in Lemma \ref{lemXY},  the Hamilton-Jacobi-Bellman (HJB) equation arising  
from dynamic programming  associated to the control problem (\ref{valuefunction}) is given by 
\begin{equation} 
\label{hjb}
\begin{cases}
\mpartialts v - \sum_{\sym\in\pm} \max_{\ell \in \{0,1\}}  \jumpintensityand{\sym} \apply{\priceand{\sym}}{v} + \tradeintensityand{\sym} \apply{\tradeand{\sym}}{v}  \; = \;   0,  
\\ 
v(T,) \; = \; x+yp-\eta y^2 
\end{cases}
\end{equation}
for $(t,p,i,s,x,y)\in[0,T]\times 2\delta \mathbb Z \times\pmset \times \Rplus\times\mathbb R \times \mathbb Z$,  where
\begin{eqnarray}
\apply{\priceand{\pm}}{v}
&:=& \Delta v(t,p \pm 2\delta i,\pm i, 0,x{+L\ell(\pm ip + \delta-\eps)},y{\mp i L\ell}),  \label{defTc}   \\ 
\apply{\tradeand{\pm}}{v} 
&:=& \int \Delta v(t,p,i,s,x +  k\ell(\pm ip + \delta-\eps),y\mp ik\ell)\, \vartheta_\pm(dk,L). \label{defJc}
\end{eqnarray}
The interpretation of these operators is rather clear: 
\begin{enumerate}
\item on one hand, $\apply{\priceand{\pm}}{v}$ represents the variation of the value function when a limit order $\ell$ in the side $\pm$ of the LOB is executed due to a price jump (this occurs with intensity rate $\jumpintensityand\pm$), hence  the  totality of size $L$ is  traded by the big market orders at the unfavorable price right after jump (adverse selection);
\item on the other hand, 
$\apply{\tradeand{\pm}}{v}$ represents the variation of the value function when a limit order $\ell$ in the side $\pm$ of the LOB is executed by a small market order (which arrives with intensity $\tradeintensityand\pm$), 
hence a quantity $k$ $\in$ $(0,\ldots,L)$ is traded  with probability $\vartheta_\pm(dk,L)$ at the favorable current price. 
\end{enumerate}

 By considering the particular hold strategy, i.e. $\bell$ $=$ $0$, in  \reff{valuefunction},  we have:
\beq \label{lowerv} 
v_{hold}(t,p,i,s,x,y) \; := \;  x + y \condprob{E}{t,p,i,s}{P_T} - \eta y^2 & \leq & v(t,p,i,s,x,y), 
\enq
and we denote by  $v_{mm}$ $:=$ $v-v_{hold}$ the nonnegative function, representing  the additional value with respect to the hold-strategy, due to  
the optimal market making strategy.  From  Lemma \ref{lemma::theta}, and setting the new state variable, called \important{strong inventory}
\beqs
q &:= & iy, 
\enqs
we see that  $v_{hold}$ can be decomposed  as: 
\beqs
v_{hold}(t,p,i,s,x,y) &=& \left(x + yp\right) + \left(q\theta(t,s)\right) -  (\eta q^2)  \\
&=&  \text{portfolio value} \; +  \; \text{martingale deviation} - \; \text{inventory penalisation}
\enqs 
The first term of this decomposition is the sum of the current portfolio portfolio value, valued at the mid-price, while the second term is proportional to $q$. The function $\theta(t,s)$ (studied in paragraph \ref{sectheta})  represents the average distance from the martingale price, i.e. it measures the average behavior of the stock price in terms of drift and reversion.  For $\theta>0$ (resp. $\theta<0$), the average value of stock price at maturity will reflect a drifting (resp. reverting) component, while for $\theta\equiv 0$, the stock price is a martingale.

  The next result provides an a priori upper bound on the value function $v$.

\begin{Lemma} \label{boundv} 
There exists some positive constant $C$ s.t. 
\beqs
v(t,p,i,s,x,y) & \leq &  x + yp + e^{C(T-t)}( 1 + |y|), 
\enqs
for all  $(t,p,i,s,x,y)$ $\in$ $[0,T]\times 2\delta\Z \times \pmset \times \Rplus \times\R\times\Z$. 
\end{Lemma} 
\begin{proof}  Let us consider the function: $\varphi(t,p,x,y)$ $:=$ $x + yp + e^{C(T-t)}( 1 + |y|)$ for some positive constant $C$.  
Then, under the assumption that $h_\pm$ and $\lambda_\pm$ are bounded, say by $B_\infty$, a straightforward calculation shows that: 
\beqs
& & - \Dt{\varphi}   - \sum_{\sym\in\pm} \max_{\ell \in \{0,1\}} h_\nu(s) \apply{\priceand{\sym}}{\varphi} 
+  \tradeintensityand{\sym} \apply{\tradeand{\sym}}{\varphi} \\
& \geq & Ce^{C(T-t)}(1 + |y|) -   2 B_\infty \big[ L|\delta-\eps| + 2 \delta |y| + 2 L e^{C(T-t)}  \big] \; \geq \; 0, 
\enqs
for $C$ large enough.  By Dynkin's formula for pure jump processes, we deduce that the process $\varphi(t,P_t,X_t,Y_t)$ is a supermartingale for any 
$\bell$ $\in$ $\Ac$, and so: 
\beq \label{inegsuper}
\E_{t,p,i,s,x,y} \big[ X_T+Y_TP_T-\eta Y_T^2 \big] \; \leq \;  \E_{t,p,i,s,x,y} \big[ \varphi(T,P_T,X_T,Y_T) \big] & \leq & \varphi(t,p,x,y). 
\enq
Since $\bell$ $\in$ $\Ac$ is arbitrary in the above inequality,  this shows  the required  upper bound: $v$ $\leq$ $\varphi$. 
\end{proof}

The above lemma together with the lower bound \reff{lowerv}, and recalling that $\theta$ is bounded (see Proposition \ref{prop::theta}) shows that 
the value function $v$ is well-defined and finite, and satisfies a linear growth condition in $x$, and a quadratic growth condition in $y,p$.

In order to solve the HJB equation associated to the optimal control problem (\ref{valuefunction}), we will go through the following steps:
\begin{enumerate}
\item we  first consider the case where the  agent is not affected by any risk aversion (${\eta=0}$), and provide analytical formula for the optimal controls and a numerical approach to compute the value function;
\item thanks to a  suitable change of variable, we perform a dimension reduction of the HJB equation associated to the control problem with risk aversion $\eta$ $>$ $0$,  and obtain optimal controls as a non-explicit deformation of the previous case;
\item finally, by a perturbation method when $\eta \rightarrow 0^+$, we provide an explicit form for optimal controls with small risk aversion, in terms of the resolution of  \important{four linear PDE's}, 
and give a financial interpretation of this result. 
\end{enumerate}

\subsection{The  no risk aversion case}

In this section, we will deal with the special case where $\eta=0$. An agent with no risk aversion does not worry about her market position since
neither any penalty for holding a large inventory affects her utility function, nor she has any inventory constraint. 
Our guess is that in the decomposition of the value function: $v$ $=$ $v_{hold}$ $+$ $v_{mm}$,  the market making part $v_{mm}$, is 
completely independent from the current portfolio, i.e. from the variables $(x,p,i,y)$, and would depend only on $(t,s)$ in the no-risk aversion case.

 We  have indeed  the following characterization of the value function and optimal controls.

\begin{Theorem}\label{prop::eta0}
For $\eta=0$, the value function of the control problem (\ref{valuefunction}) is given by 
\begin{eqnarray*}
v &=& v_{hold} + v_{mm}, \\
v_{hold}(t,p,i,s,x,y) &=& x + yp + q \theta(t,s),\\
v_{mm}(t,s) &=& \omega(t,s), 
\end{eqnarray*}
where 
\beq \label{repreomega}
\omega(t,s) &=&  \sum_{\sym\in\pm} \condprob{E}{t,s}{\int_t^T \maxx{\barrier_\sym(u,S_u)} du}, 
\enq
for $(t,s)\in[0,T]\times\Rplus$, with
\begin{eqnarray} \labeleq{Fpm}
\barrier_\pm(t,s) &:= & {\tradeintensityand{\pm}\left(\delta-\eps\mp\theta(t,s)\right) \vartheta_\pm^1(L)}  
  - \;   {\jumpintensityand{\pm}\left(\delta+\eps+\thetatzero\right) L}, 
\end{eqnarray}
(recall $\vartheta_\pm^1(L)$ in \reff{momentvartheta}).  Moreover, the optimal controls are given in feedback form by:
\beq \label{optimal_control_0} 
\hat \ell_\pm(t,s) &=& \indicator{\barrier_\pm(t,s) > 0}\for\;\;  t \in [0,T], s \in \Rplus.
\enq
\end{Theorem}

\begin{proof}
The problem \reff{valuefunction} lies in the general class of Markov decision processes (see the recent book  \cite{baurie11}), and it is known (see e.g. \cite{son86} or \cite{davfar99}) that the corresponding value function $v$ is a  viscosity solution  
to the HJB equation \reff{hjb}. Let us now consider as a candidate $u$ for $v$ a function in the form: $u(t,p,i,s,x,y)$ $:=$ 
$x + yp + q \theta(t,s) + \omega(t,s)$, for some function $\omega$ to be determined. 
First, observe that  $u$ satisfies the terminal condition in \reff{hjb} for $\eta$ $=$ $0$:   $u(T,p,i,s,x,y)$ $=$ $x+yp$ iff  $\omega$ satisfies the terminal condition: $\omega(T,s)$ $=$ $0$. Moreover,  
by definitions \reff{defTc} and \reff{defJc} of the operators  $\mathcal J_\pm$ and  $\Tc_\pm$, we have 
\begin{eqnarray*}
\partialts u &=& q{\left(\partialts\theta\right)} +\left(\partialts\omega\right),\\
\apply{\priceand\pm}{u}
&=&  q {\left( {\pm 2 \delta  \pm \thetatzero-\theta(t,s)}\right)} + \omegadiff +L\advsel , \\
\apply{\tradeand\pm}{u}
&=& \int k \ell  \left(\delta-\eps \mp \theta(t,s)\right) \vartheta_\pm(dk,L).
\end{eqnarray*}
We then see that $u$ is solution to the HJB equation \reff{hjb} iff $\omega$ satisfies: 
\begin{equation}  \labeleq{omega}
\begin{cases}
- \Dt{\omega} - \Ds{\omega} - \vol\Delta\omega(t,0)  
- \sum_{\sym\in\pm} \maxx{\barrierand\sym} +  \underbrace{q\left(- {\cal M}\theta -2\delta\drift\right)}_{=0 \text{ by }  \refequation{theta}} \; = \;  0, \\
\omega(T,)  \; = \;  0, 
\end{cases} 
\end{equation}
Notice  that $G_\pm$ is bounded as $h_\pm$ and $\lambda_\pm$ are assumed to be bounded. 
Then, the (nonnegative) function defined by: 
\beqs
\tilde\omega(t,s) & := &  \sum_{\sym\in\pm} \condprob{E}{t,s}{\int_t^T \maxx{\barrier_\sym(u,S_u)} du}, 
\enqs
is  bounded, and is a viscosity solution to \refequation{omega} by  Feynman-Kac representation for such  linear integro-differential equation,  recalling that  $(S_t)$ is a piecewise deterministic jump process with intensity 
$\sigma^2(S_t)$.  Therefore, by taking $\omega$ $=$ $\tilde\omega$,  we see by construction that the function $u(t,p,i,s,x,y)$ $:=$  $x + yp + q \theta(t,s) + \omega(t,s)$ is a viscosity solution to  the HJB equation \reff{hjb}. By uniqueness for such first-order integro-differential equation \reff{hjb}  (see \cite{say91}), we deduce  that $v$ $=$ $u$.

 Next, let us consider the feedback control $\hat\ell_\pm$ on the side $\pm$ of the LOB obtained from the argument maximum over $\ell$ $\in$ $\{0,1\}$ of
$$
\jumpintensityand\pm  \apply{\priceand{\pm}}{v}  +   \lambda_{\pm} \apply{\tradeand{\pm}}{v}.
$$
In other words, 
$$
\hat\ell_\pm = 1 \; \Longleftrightarrow \; \jumpintensityand\pm \langle \mathcal J_\pm[1],v \rangle   
+    \tradeintensityand\pm \langle \Tc_\pm[1],v \rangle   \; > \; 
\jumpintensityand\pm \langle \mathcal J_\pm[0],v \rangle   +   \tradeintensityand\pm \langle \Tc_\pm[0],v \rangle.
$$
By substituting the decomposition $v$ $=$  $x + yp + q \theta(t,s) + \omega(t,s)$,  we see that  $\hat\ell_\pm$ $=$ $\hat\ell_\pm(t,s)$ and 
$$
\hat\ell_\pm(t,s) \; = \; 1 \; \Longleftrightarrow \;  \barrier_\pm(t,s)  \; > \; 0. 
$$
Let us finally check that such feedback control $\hat\ell_\pm$ provides an optimal control. Actually,  since it attains the argument maximum in the HJB equation, we have 
\begin{equation} 
\label{hjbopt}
\begin{cases}
\mpartialts v - \sum_{\sym\in\pm}    \jumpintensityand{\sym} \langle\price_\nu[\hat\ell_\nu], v\rangle + \tradeintensityand{\sym} \langle\trade_{\sym}[\hat\ell_\nu],v\rangle  \; = \;   0,  
\\ 
v(T,)  \; = \;  x+yp,  
\end{cases}
\end{equation}
so that by Feynman-Kac representation for the above linear integro-differential equation: 
\beqs
v(t,p,i,s,x,y) &=&  \condprob{E}{t,p,i,s,x,y}{\hat X_T +  \hat Y_T P_T},
\enqs
where $\hat X$ and $\hat Y$ are the wealth and inventory process with dynamics in Lemma \reff{lemXY}, controlled by the strategy $(\hat\bell_t)_t$ $=$ 
$(\hat\ell_+(t,S_{t^-}),\hat\ell_-(t,S_{t^-}))_t$ $\in$ $\Ac$. This  shows the optimality of this feedback control. 
\end{proof}

\subsubsection*{Financial interpretation}

\noindent  The strong (resp. weak) optimal policies described in \reff{optimal_control_0} are binary controls depending on $\barrier_\pm(t,s)$. 
Cutting the surfaces $z=\barrier_\pm(t,s)$ with the hyperplane $z=0$ and projecting on $(t,s)$-plane the portions above the hyperplane, we obtain the trading regions. 
Since controls drive execution, and execution is due either on a trade or a price jump, optimal controls are naturally decomposed into two parts, as shown by the expression of $\barrier_\pm$ in  \refequation{Fpm}: 
\beqs
\barrier_\pm(t,s) &=& G^{\text{trd}}_\pm(t,s) - G^{\text{jmp}}_\pm(t,s), \nn \\
&:=& \tradeintensityand\pm\left(\delta-\eps\mp\theta(t,s)\right) \vartheta_\pm^1(L) - \jumpintensityand\pm\left(\delta+\eps+\thetatzero\right) L \nn. 
\enqs
\begin{enumerate}
\item The \important{trade part} $G^{\text{trd}}_\pm$: the agent orders are matched by small market orders (not impacting the stock price) of random size. For each executed lot she gains the half-spread thanks to passive execution (market making gain), she loses the transaction costs, and since her inventory changes, she loses the martingale distance $\theta(t,s)$ that she would gain (in average) keeping her position until the horizon (attention: this quantity might be negative).
This scenario is favourable for the agent, since she has the time to close her spread before the stock price jumps again and she has gained half a tick w.r.t. to the mid-price evaluation of the stock: this profit comes from small uniformed traders, that are unable to make the stock price jump and need the liquidity provided by the market maker.
\item  The \important{jump part} $G^{\text{jmp}}_\pm$: the agent orders are cleared 
at rate $\jumpintensity_\pm(S_t)$ by big market orders impacting the stock price. As before, for each executed lot she gains the half-spread thanks to passive execution (market making gain), but loses a spread because of the stock price jumps, ending with a passive of half a spread, she loses the transaction costs, and since her inventory changes, she loses the martingale distance $\theta(t,0)$ (the price has just jumped). 
Looking at \refequation{Fpm}, it is clear that the limit order execution due to  big orders is a source of risk for the agent, called \important{adverse selection}. This is due to the fact that large market orders makes the price jump, and thus do not allow  the agent to close her spread. On the contrary, the agent finds herself buying/selling $L$  lots (the maximum admissible size) when the stock price is suddenly decreasing/increasing value. 
\end{enumerate}


As one can see, the  ``trade"  case  represents a positive event for the agent, since it leads to a proper market making strategy collecting spreads where buy and hold portfolio would not lead to substantial gains, while the ``jump"  represents a negative event, since it leads to an unfavourable execution, where we sell (resp. buy) a stock whose value has decreased (resp. increased). 
Hence, our  marked point process modeling of  stock price leads to a much more realistic model, evaluating the risk of being executed under negative circumstances, which would not be possible if the stock price were a continuous process.  The quantity $\barrier_\pm$  is thus interpreted as the portfolio value arising from market making strategy, and naturally the  value function $v_{mm}$ $=$ $\omega$ in \reff{repreomega} 
is the expected gain of the portfolio value.
Optimal controls are determined by a trade-off among
\begin{enumerate}
\item pure market making gain due to round-trip  executions,
\item trend or reversion anticipation thanks to the information associated to $\theta(t,s)$, 
\item limitation of  adverse selection risk.
\end{enumerate}
By stressing the dependence of optimal control on horizon $T$, we write  $\hat\ell^T_\pm(t,s)$ for  $\hat\ell_\pm(t,s)$, and we see from \reff{optimal_control_0} and \refequation{theta_infty} that it converges for large horizon to the stationary value:
\beq \label{optimal_control_0_infty}
\lim_{T\rightarrow\infty} \hat\ell^T_\pm(t,s) &=& \indicator{G^\infty_\pm(s) > 0}
\enq
with  
\beqs
G^\infty_\pm(s) &:=&  \tradeintensityand\pm\left(\delta-\eps\mp\theta^\infty(s)\right) \vartheta_\pm^1(L) 
\;  -  \; \jumpintensityand\pm\left(\delta+\eps+\theta^\infty(0)\right) L\nn. 
\enqs
We plot  in Figure \ref{fig::omega} the market making performance $\omega$, and in Figure \ref{fig::no_risk_aversion} the optimal limit order controls.

\newfigure{omega}
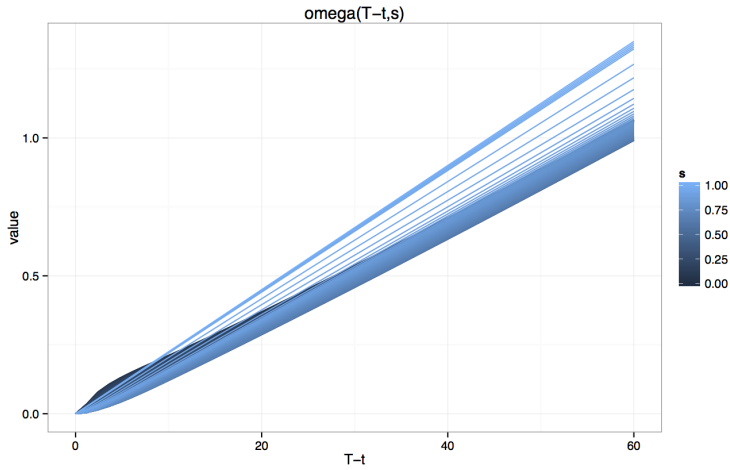
{$\omega(T-t,s)$ for diffe\-rent s-sections. Far from the horizon, the agent has a linear average gain per time unit depending on the starting elapsed time. In the distribution queue, the agent  gains more since she enters the market at a stable time, where the stock price is not likely to jump, representing a perfect scenario for the market maker.\ep}

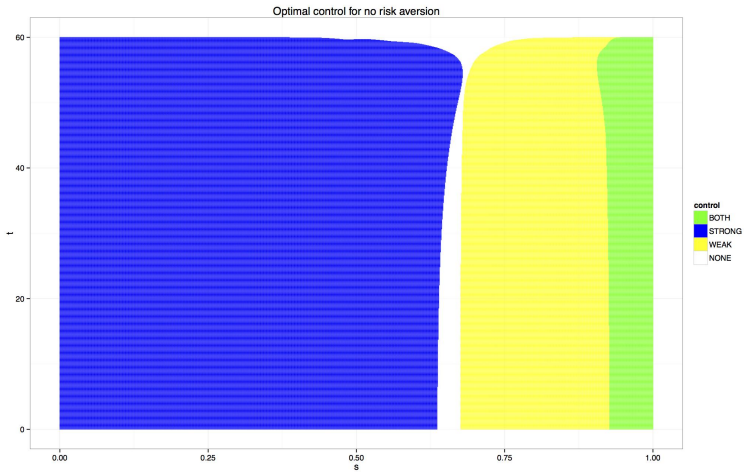
\newfigure{no_risk_aversion}{optimal policy for $\eta=0$. 
The agent places her limit order on the strong side for both small and big elapsed times, while posting on the weak side is restricted to large $s$: this is due to behavior of $\theta(t,s)$, changing signs in the right side of the cart, inducing the maker maker to place orders on the side to exploit the anticipation on the next jump. For $t\rightarrow 0$, the horizon is far and the policy stabilizes to her asymptotic value described in \reff{optimal_control_0_infty}\ep.
}


\subsection{The small risk aversion case}

In this section, we  prove that for small risk aversions, the solution of the HJB equation (\ref{hjb}) associated to the optimal control problem (\ref{valuefunction}) is a deformation of the solution in  the case $\eta=0$, and we explicitly characterize this deformation.   We first give  the exact solution of the problem in terms of a non-explicit deformation of the 
$(\eta=0)$-value function, that one can evaluate numerically if not wanting to involve approximation arguments. This result will help us to illustrate, in terms of probabilistic representation,  how a particular strategy affects the value function.

\begin{Theorem}\label{thm::eta0}
The value function associated to the control problem (\ref{valuefunction}) is given by
\begin{eqnarray}
v(\allvar) &:=& v^{(\eta)}(\allvar) \; = \;  v^{(0)}(\allvar) -\Risk(t,s,q), \label{value_function}
\end{eqnarray}
where $v^{(0)}(\allvar)$
is the solution of the control problem in the no risk aversion case  (Theorem  \ref{prop::eta0}), while $\Risk(t,s,q)$ is nonnegative and is  the  unique viscosity solution with quadratic growth in $q$  to 
\begin{equation} \label{hjb_riskaversion} 
\begin{cases}
 \mpartialts{\Risk} -  \sum_{\sym\in\pm} 
   \min_{\ell \in \{0,1\}}  \Big[  \apply{\mathcal R_\sym[\ell]}{\Risk}  +   \maxx{\barrierand\sym} -\ell \barrierand\sym \Big]  \; =  \;  0, \\
\Risk(T,) \; = \;  \eta q^2,
\end{cases}
\end{equation}
for $(t,s,q) \in [0,T]\times \Rplus \times \mathbb Z$, where $\barrierand\sym$ is given by \refequation{Fpm} and
\begin{eqnarray}
\apply{\mathcal R_\pm[\ell]}{\Risk} & := &  \jumpintensitypm \Delta\Risk(t,0,\pm q-L\ell) \label{Rpm} + \tradeintensitypm  \int \Delta\Risk(t,s,q \mp k\ell)  \law \nn.  
\end{eqnarray}
Moreover, the optimal controls for problem (\ref{valuefunction}) are given in feedback form by
\begin{eqnarray} \label{optimaleta}
\hat \ell_\pm(t,s)  \equaldef \hat \ell_\pm^{(\eta)}(t,s) &=&   \indicator{\barrier_\pm(t,s) > \apply{\mathcal C_\pm}{\Risk}}, 
\end{eqnarray}
where
\begin{eqnarray}
\apply{\mathcal C_\pm}{\Risk} := \jumpintensitypm \left(\Risk(t,0,\pm q-L)-\Risk(t,0,\pm q)\right) +   \tradeintensitypm\int \Delta\Risk(t,s,q \mp k) \law\nn\stop 
\end{eqnarray}
\end{Theorem}

\begin{proof} Let us first derive formally the equation that should be satisfied by $\Risk$ when using the ansatz  \reff{value_function}.   
By definitions of the operators in \reff{defTc} and \reff{defJc}, we have
\begin{eqnarray*}
\Dt{v^{(\eta)}}  + \Ds{v^{(\eta)}}
&=& \left(\partialts{v^{(0)}}\right)  - \left(\partialts{\Risk}\right), \labeleq{partialts1}\\
\apply{\priceand{\pm}}{v^{(\eta)}} 
&=& \apply{\priceand{\pm}}{v^{(0)}}  - \Delta\Risk(t,0,\pm q-L\ell), \\
\apply{\tradeand{\pm}}{v^{(\eta)}}  
&=& \apply{\tradeand{\pm}}{v^{(0)}}  - \int\Delta\Risk(t,s,q \mp  k\ell)  \law.
\end{eqnarray*}
Plugging into \reff{hjb}, and using the decomposition of $v^{(0)}$ in Theorem \ref{prop::eta0}, we see after some straightforward calculations that  $\Risk$ should satisfy the PDE \reff{hjb_riskaversion}.  
Now, let us prove that $v^{(\eta)}$ is indeed in the form \reff{value_function}.  From the dynamics of the strong inventory process $Q$ defined in Remark 
\ref{remQ}, we notice that equation \reff{hjb_riskaversion} is actually  the HJB equation associated to the control problem: 
\beq \label{tildezeta}
\tilde\zeta^{(\eta)} (t,s,q) &:=& \inf_{\bell\in\Ac} \E_{t,s,q} \Big[ \int_t^T \sum_{\nu\in\pm}  \big[\max(G_\nu(u,S_u),0) - \ell_u^\nu G_\nu(u,S_u) \big] du 
+ \eta Q_T^2 \Big],
\enq  
for $(t,s,q)$ $\in$ $[0,T]\times\R_+\times\Z$. It is clear that $\tilde\zeta^{(\eta)}$ is nonnegative, and by taking the zero control in \reff{tildezeta}, we see that $\tilde\zeta^{(\eta)}(t,s,q)$ $\leq$ $\omega(t,s)$ $+$ $\eta q^2$, recalling the expression \reff{repreomega} of $\omega$.  Since $\omega$ is bounded, this shows that $\tilde\zeta^{(\eta)}$ is of quadratic growth in $q$, uniformly in $(t,s)$. We then know from \cite{son86} and  \cite{say91} that 
$\tilde\zeta^{(\eta)}$ is the unique viscosity solution to \reff{hjb_riskaversion}. Therefore, by defining the function $\zeta^{(\eta)}$ $=$ $\tilde\zeta^{(\eta)}$, 
we see by construction that the function $u^{(\eta)}$ $:=$ $v^{(0)}$  $+$ $\zeta^{(\eta)}$ is a viscosity solution to the HJB equation \reff{hjb}, and by uniqueness for this  equation (see \cite{say91}), we deduce that $v^{(\eta)}$ $=$ $u^{(\eta)}$.

 Let us consider the feedback controls $\hat\ell_\pm^{(\eta)}$, for both strong and the weak side, that attain the maximum between $\apply{\mathcal R_\pm[0]}{\Risk}$ and  
$G_\pm(t,s)$ $-$ $\apply{\mathcal R_\pm[1]}{\Risk}$, and given precisely in  the form \reff{optimaleta}. Then, from the decomposition \reff{value_function} of $v^{\eta)}$,  $\hat\ell_\pm^{(\eta)}$ attains actually the maximum over 
$\ell$ $\in$ $\{0,1\}$ of:
$$
\jumpintensityand\pm  \apply{\priceand{\pm}}{v^{(\eta)}}  +   \lambda_{\pm} \apply{\tradeand{\pm}}{v^{(\eta)}},
$$
and by same argument as in Theorem \ref{prop::eta0} for $\hat\ell_\pm^{(0)}$,  this shows the optimality of $\hat\ell_\pm^{(\eta)}$. 
\end{proof}

The deformation function $\Risk$ due to risk aversion is solution to the non linear integro-differential equation \reff{hjb_riskaversion}, which can be solved numerically.  We use instead a perturbation approach  for  deriving a first-order expansion of $\Risk$ for small risk aversion $\eta$. 


\begin{Theorem} \label{thm::asymptotic}
The function  $\Risk(t,s,q)$ can be linearly approximated in $\eta>0$ by 
\begin{eqnarray} \label{approx}
\Risk(t,s,q) & = & \eta\left(q^2 +    2q \risk_1(t,s) +  \risk_0(t,s) \right) - R^{(\eta)}(t,s,q),
\end{eqnarray}
where 
\begin{enumerate}
\item $\risk_1$ is the unique bounded continuous viscosity solution to the linear integro-differential equation: 
\begin{equation}   \label{eqzeta1}
\begin{cases}
- {\cal M}\risk_1 
+ \sum_{\sym\in\pm} \sym  			\left( \jumpintensity_\sym(s)L  + \tradeintensity_\sym(s)\vartheta_\nu^1(L)\right) \indicator{\barrierand\sym>0} = 0  \\	
\risk_1(T,) = 0 
\end{cases}
\end{equation} 
with $\mathcal M$  the linear operator defined in \refequation{theta}, and admits the probabilistic representation:
\beqs
\risk_1(t,s) &=& \E\big[ \hat Y_T | \hat Y_t=0,I_t=+1,S_t=s],
\enqs
where $\hat Y$ is the inventory process controlled by the feedback strategy $\hat\bell$ $=$  $(\hat\ell_+(t,S_{t^-}),\hat\ell_-(t,S_{t^-}))_t$ defined  in \reff{optimal_control_0},  
\item  
\beq
\risk_0(t,s) &=& \sumpm \E_{t,s} \Big[  \int_t^T  \sum_{\sym\in\pm}  \big[  \jumpintensity_\sym(S_u) \big(L^2 -  2L \risk_1(t,0) \big)  \label{repzeta0} \\
& & \hspace{2cm}   + \;  \tradeintensity_\sym(S_u)\big(\vartheta^2_\sym(L) - 2\sym\vartheta^1_\sym(L)\risk_1(u,S_u)\big) \big]  \indicator{G_\nu(u,S_u)>0} \, du \Big]  \nonumber 
\enq
is the unique bounded  continuous  viscosity solution of 
\begin{equation}  \label{eqzeta0}
\begin{cases}
\mpartialts{\risk_0} - \vol\Delta\zeta_0(t,0)    \\
- \sum_{\sym\in\pm}  \big[  \jumpintensity_\sym(s) \big(L^2 - 2L \risk_1(t,0)\big) + \tradeintensity_\sym(s)\big(\vartheta^2_\sym(L) - 2\sym\vartheta^1_\sym(L)\risk_1(t,s)\big) \big] 
\indicator{\barrierand\sym>0} \; =  \; 0,  \\
 \risk_0(T,) \; =\;  0,
\end{cases}
\end{equation}
\item the remainder $R^{(\eta)}$ is a non-negative function, s.t.  $R^{(\eta)} = o(\eta)$, i.e.  
$$
\lim_{\eta\rightarrow 0^+} \eta^{-1}R^{(\eta)}(t,s,q) = 0 \for \forall(t,s,q) \in [0,T]\times\R_+\times\Z.
$$. 
\end{enumerate}
\end{Theorem}
 
\begin{proof}  
\underline{{\it  Step 1.}} Recalling the dynamics of $\hat Y$ in Lemma \ref{lemXY}, we notice that the triple $(\hat Y_t,I_t,S_t)$ is a Markov process, and let us consider the function: 
\beqs
\hat\Yc(t,y,i,s) & := &  \E_{t,y,i,s}[\hat Y_T]. 
\enqs
From standard result on pure jump process (see e.g. \cite{baurie11}), and under continuity assumptions on the intensity functions $h_\pm$ and $\lambda_\pm$, the function $\hat\Yc$ is continuous, and by the Feynman-Kac representation, it is solution to the linear integro-differential equation: 
\begin{equation} \label{eqYc}
\begin{cases}
\mpartialts{\hat\Yc} \;   \;  - \sumpm h_\sym(s) \Delta \hat\Yc(t,  y - \sym i  L \hat\ell_{\sym}(t,s), \sym i,s) \\
\; - \;  \sumpm \lambda_\sym(s) \int \Delta \hat\Yc(t, y - \sym i  k \hat\ell_{\sym}(t,s), i,s ) \vartheta_{\sym}(dk, L) \;  = \;  0,  \\
\hat\Yc(T,) \; = \;  y.
\end{cases}
\end{equation}
Then, by the same arguments as in Proposition \ref{lemma::theta}, one checks  that $\hat\Yc$ is decomposed into: 
\beqs
\hat\Yc(t,y,i,s) &=& y + i \risk_1(t,s),
\enqs
where $\risk_1$ is the unique bounded continuous viscosity solution to \reff{eqzeta1}, and also have the announced probabilistic representation since  $\risk_1(t,s)$ $=$ $\hat\Yc(t,0,1,s)$.

\underline{{\it  Step 2.}}  Let us define the function $\test$ $=$ $\test(t,s,q)$  by $\test := \Risk/\eta$, which satisfies from  (\ref{hjb_riskaversion}): 
\begin{equation}
\begin{cases}
\eta\left(\mpartialts{\test}\right)  \nonumber \\ 
\; - \;  \sum_{\sym \in \pm} \min_{\ell \in \{0,1\}}  \left[  \eta \apply{\mathcal R_\sym[\ell]}{\test}  + \maxx{\barrierand\sym} - \ell \barrierand\sym \right]  \; = \;  0,  \nonumber \\
\test(T,s,q)  \; = \;   q^2,
\end{cases}
\end{equation}
and notice that  $\mathcal R_\pm$ defined in \reff{Rpm} admits an affine decomposition $\mathcal{R}_{\pm}[\ell] = \mathcal{R}_{\pm}^{0} + \ell \mathcal{R}_{\pm}^{1}$, where
\begin{eqnarray}
\apply{\mathcal R_\pm^0}{\test} & := & \jumpintensitypm  \Delta \test(t,\pm q, 0), \nn \\
\apply{\mathcal R_\pm^1}{\test}  & := & \jumpintensitypm \round{\test(t,\pm q - L, 0) - \test(t,\pm q, 0)}  + \tradeintensitypm \int \Delta \test(t, q \mp k,s) \vartheta_{\pm}(dk, L). \nn
\end{eqnarray}
We can then rearrange the equation satisfied by  $\test$ as
\begin{equation}
\begin{cases}
\eta\left(\mpartialts{\test}\right)  -  \eta \sum_{\sym \in \pm} \apply{\mathcal R_\sym^0}{\test} \\
\; + \;   \sum_{\sym \in \pm} \big[   \maxx{\barrierand\sym - \eta \apply{\mathcal R_\sym^1}{\test}} - \maxx{\barrierand\sym} \big] \; = \; 0 \nn \\
\test(T,s,q)  \; = \;   q^2. 
\end{cases}
\end{equation}
Now, by observing that 
\begin{eqnarray*}
\maxx{x-\varepsilon} -  \maxx{x} &=&  - \varepsilon \indicator{x>0} + z_\eps(x),
\end{eqnarray*}
where $z_\eps(x)$ is a non-negative function  satisfying:  $z_\eps(x)$ $\leq$ $\vert \eps \vert \indicator{\vert x \vert < \vert \eps \vert}$,  we can write the PDE for $\test$ as
\begin{equation} \label{eqtest}
\begin{cases}
- \Lc\test + Z^{(\eta)}(t,s,q)  \; =  \;  0, \\
\test(T,s,q) \; = \; q^2
\end{cases}
\end{equation}
where $\Lc$ is the linear operator:
\beqs
\Lc\test & := & \Dt{\test} + \Ds{\test} +  \sum_{\sym \in \pm} \apply{\mathcal R_\sym^0}{\test} +  \sum_{\sym \in \pm}  \apply{\mathcal R_\sym^1}{\test}\indicator{\barrierand\sym > 0}  \\
& = & \Dt{\test} + \Ds{\test} +  \sum_{\sym \in \pm} \apply{\mathcal R_\sym[\hat\ell_\sym]}{\test}, 
\enqs
which is actually the infinitesimal generator of the Markov process $(t,S_t,\hat Q_t)$ where $\hat Q_t$ $=$ $I_t\hat Y_t$ is the strong inventory process controlled by $\hat\bell$, 
and  
\begin{eqnarray}
0 \; \leq \; Z^{(\eta)}(t,s,q) & := & \frac{1}{\eta} \sumpm \big[ \maxx{G_\sym(t,s) - \eta \apply{\mathcal R_\sym^1}{\test}}-\maxx{G_\sym(t,s)} +  \eta \apply{\mathcal R_\sym^1}{\test} 
1_{G_\sym(t,s) > 0 } \big] \nonumber \\
  &\leq & \sumpm \abs{\apply{\mathcal R_\sym^1}{\test}} \indicator{\abs{G_\sym(t,s)} < \eta \abs{\apply{\mathcal R_\sym^1}{\test}}} =:  \overline Z^{(\eta)}(t,s,q) \; \rightarrow \; 0, \; \mbox{ as } \eta \; \mbox{goes to zero}. 
  \label{Zeta}
\end{eqnarray}

\underline{{\it  Step 3.}}  We now approximate $\test$ by the solution to the linear parabolic integro-differential equation: 
\begin{equation} \label{eqtildetest}
\begin{cases}
- \Lc\tilde\test   \; =  \;  0, \\
\tilde\test(T,s,q) \; = \; q^2,
\end{cases}
\end{equation}
which is represented the Feynman-Kac formula: 
\beqs
\tilde\test(t,s,q) &=& \E_{t,s,q} \big[ \hat Q_T^2]. 
\enqs
Actually, by considering the function 
\begin{eqnarray*}
\tilde \Phi(t, s, q) & := & q^2 + 2 q  \zeta_1(t,s) + \zeta_0(t,s),
\end{eqnarray*}
where $\risk_1$ is defined from Step 1,  and $\risk_0$ defined in  \reff{repzeta0},  we see after some tedious but straightforward calculation that 
\beqs
\Lc\tilde\Phi &=&  2q \Big(  - \mathcal M \zeta_1 
+  \sumpm \sym  \round{h_\sym(s) +  \lambda_\sym(s) \vartheta^1_\sym(L)} \indicator{\barrierand\sym > 0} \Big) +  \\
& &   \Dt{\risk_0} + \Ds{\risk_0} +  \vol\Delta\zeta_0(t,0)   +  \sum_{\sym\in\pm}  \jumpintensity_\sym(s) \big(L^2 - 2L \risk_1(t,0)\big) + \tradeintensity_\sym(s)\big(\vartheta^2_\sym(L) - 2\sym\vartheta^1_\sym(L)\risk_1(t,s)\big).
\enqs
Since $\zeta_1$ satisfies equation \reff{eqzeta1},  and  $\zeta_0$ is the unique bounded solution to \reff{eqzeta0} by Feynman-Kac representation, this shows that  $\tilde\Phi$ is solution to \reff{eqtildetest}, and by uniqueness, we deduce that 
$\tilde\test$ $=$ $\tilde\Phi$.  Finally, it remains to prove that $R^{(\eta)}$ $:=$ $\eta\tilde\Phi-\zeta^{(\eta)}$ $=$ $\eta(\tilde\test - \test)$ is a nonnegative  $o(\eta)$ function, i.e. $H$ $:=$ $\tilde\test-\test$ is a nonnegative function converging to zero as $\eta$ goes to zero. 
From \reff{eqtest} and \reff{eqtildetest}, and since  $\Lc$ is a  linear operator,  the function $H$ is solution to: 
\begin{equation} 
\begin{cases}
- \Lc H + Z^{(\eta)}(t,s,q)  \; =  \;  0,  \nn \\
R^{(\eta)}(T,s,q) \; = \; 0,
\end{cases}
\end{equation}
hence given by the Feynman-Kac formula:  
\begin{eqnarray*}
H(t,s,q) &=&  \E_{t,s,q}\big[ \int_t^T  Z^{(\eta)}(u, S_u, \hat Q_u) du \big].  
\end{eqnarray*}
By \reff{Zeta} and monotone convergence theorem,  we conclude that $H$ is nonnegative, and converges to zero as $\eta$ goes to zero. 
\end{proof}



\begin{Remark}
{\rm The computation of $\Risk(t,s,q)$ through the equation \reff{hjb_riskaversion} requires the numerical resolution of a {non-linear system} of $1$-dimensional PDE indexed by $q\in\mathbb Z$. 
Alternatively, the first-order expansion for small risk aversion, with a quadratic in $q$  leading term, involves the computation of  $(\risk_0,\risk_1)$ through a system of linear PDE's.  
It is worth noticing that: 
\begin{enumerate}
\item thanks to the approximation methods, we are led to solve four linear simple PDE's (for  $\theta,\omega,\risk_1,\risk_0$), reducing  of one the dimension of the problem;
\item since $\mathbb Z$ is an infinite set, a numerical solution needs a domain truncation and the spe\-cification of boundary conditions, while this problem does not affect the approximated solution;
\item the numerical scheme can be trivially parallelized: the solution of $\omega(t,s)$ depends on $\theta(t,s)$, while the one of $\risk_0(t,s)$ depends on $\risk_1(t,s)$, but the two couples are independent, leading to a natural parallelization.
\end{enumerate}
\ep
}
\end{Remark}

\begin{Remark}
{\rm From the structure \reff{optimaleta} of optimal control  and  the small expansion \reff{approx}, we get an approximate optimal limit order control given by:
\beqs
\tilde\ell_\pm^{(\eta)}(t,s) &:=&   \indicator{\barrier_\pm(t,s) > \apply{\mathcal C_\pm}{\tilde\zeta^{(\eta)}}} \for \text{where}\\
\tilde\zeta^{(\eta)}(t,s,q) &:=& \eta\left(q^2 + 2q\zeta_1(t,s)+ \zeta_0(t,s)\right).
\enqs
The approximate optimal feedback control can be rewritten as:
\begin{eqnarray*}
\tilde\ell_\pm^{(\eta)}(t,s) &:=&   \indicator{\barrier_\pm(t,s) > \eta \, \left(\mathbf A(t,s) \mp 2\mathbf B(t,s) q\right)}, 
\end{eqnarray*}
where
\begin{eqnarray}
\mathbf A(t,s) &:=& {\jumpintensitypm L  \left(L - 2 \risk_1(t,0)\right) + \tradeintensitypm \left(\vartheta_\pm^2(L) \mp 2\risk_1(t,s)\vartheta_\pm^1(L)\right)} \labeleq{inventory_independent} \\
\mathbf B(t,s) &:=& \jumpintensitypm L + \tradeintensitypm\vartheta_\pm^1(L) \geq 0 \labeleq{inventory_dependent}
\end{eqnarray}
$\tilde\ell_\pm^{(\eta)}$ is an approximate control of the optimal one $\hat\ell_\pm^{(\eta)}$ in the sense that  $\tilde\ell_\pm^{(\eta)}$ $-$ $\hat\ell_\pm^{(\eta)}$ $ \rightarrow$ $0$ as $\eta$ goes to zero,   since $\tilde\zeta^{(\eta)}-\zeta^{(\eta)}$ $=$  $o(\eta)$.  The control adjustment due to risk aversion can be understood as follows.
\begin{enumerate}
\item The \important{inventory independent part} in (\ref{eq::inventory_independent}) affects the agent strategy even when her inventory is ${q=y=0}$: in the no risk aversion case, the agent places a limit order if $\barrierand\sym>0$, but when $\eta>0$, this trading barrier - depending on the couple $(t,s)$ - is raised by a multiple of the risk aversion itself. The agent becomes cautious and decides to enter the market only in the most profitable cases, but  renounces to trade if expected gain is small, even if positive. This result is shown by Figure  \ref{fig::risk_aversion_control}, where the non trading region for $q=0$ grows w.r.t. the case $\eta=0$, reflecting the increased trading aversion of the agent.
\item The \important{linear part in the inventory} in (\ref{eq::inventory_dependent}) is in charge of controlling the absolute value of the inventory, in order to reduce the exposure of the agent to inventory and market risk. For $q>0$ (i.e. $iy>0$), the adjustment of the strong side (resp. weak side) turns the trading barrier down (resp. up) by a multiple of $\eta$: the agent decides to trade on the strong (resp. weak) side under less (resp. more) restrictive conditions in order to reduce (resp. not to increase) her absolute inventory. This result is summarized by Figure \ref{fig::stationary_control}, where for large values of the inventory the agent plays, independently on $s$, on one side only, in order to revert the absolute position to a smaller value.
\end{enumerate}
\ep
}
\end{Remark}

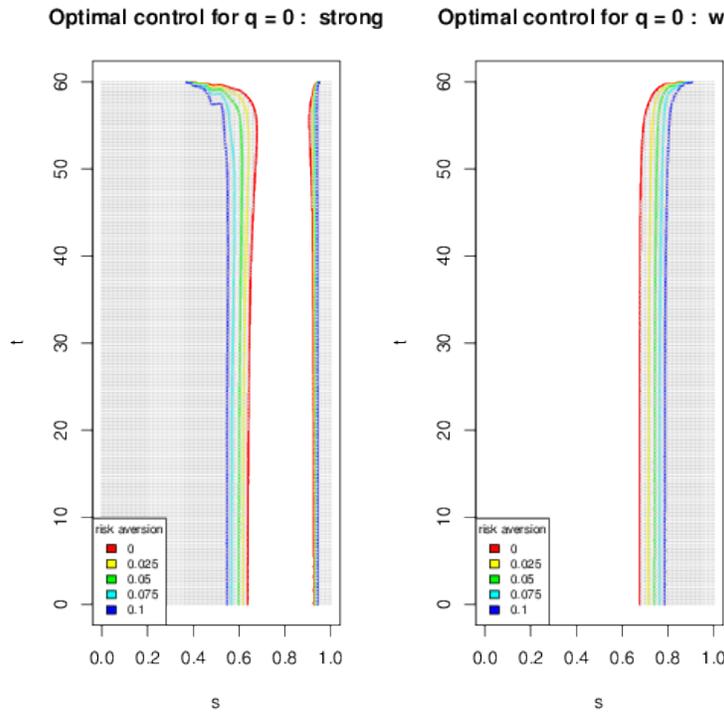
\newfigure{risk_aversion_control}{trading region for $q=0$ and different risk aversions. The chart shows how increasing the risk aversion deforms the trading region, widening the zone where the agent does not send any order.\ep}

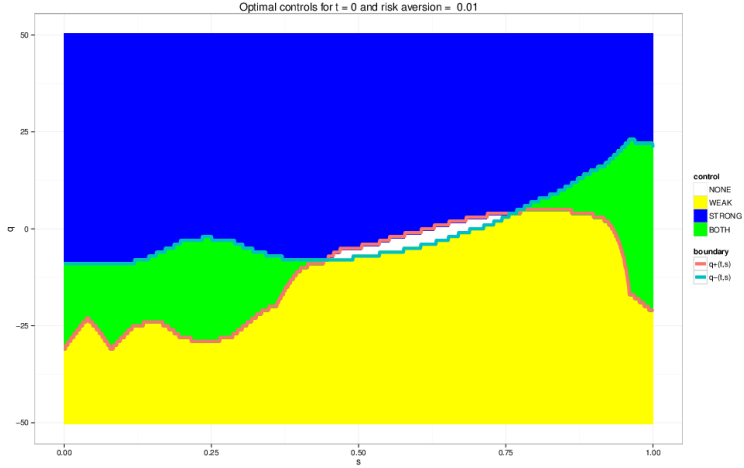
\newfigure{stationary_control}{the optimal policy at time $0$ for different values of $q$ and $s$ and positive risk aversion. The agent plays only only on the strong (resp. weak) side for large (small values) of the strong inventory in order not to be exposed to the inventory risk. For $q\approx 0$, the policy takes into account the inventory risk and the pure gain due to a market making strategy.\ep}

\newpage

\section{Conclusion and further developments}

In this paper we have exploited the framework described in the companion paper \cite{FodPha13a} in order to provide an application to a market making problem. Thanks to a Cox model, we are able to include the matching engine and complete the order book model of the best levels: only small trades are described directly, while big market orders affecting the stock price are included in the stock price dynamics itself. The perturbation technique adopted to derive optimal controls as a deformation of the no risk aversion case has helped us to improve financial interpretability and reduce the dimension problem as well as the computational cost of its numerical  solution. For further developments, more complicated stock price model can introduce statistical arbitrage opportunity, that we may capture by means of  optimal control techniques.

\small

 \appendix

\section{The estimation of the agent execution distribution $\law$}\label{appendix::estimation_vartheta}

Unfortunately, the agent execution cannot be estimated before playing or backtesting a zero intelligence strategy, i.e. placing continuously limit orders on both sides of the limit order book.  This  problem cannot be overcome easily: a high-frequency backtest platform is necessary to estimate the execution rate.  Assuming that the agent is always placed with one lot on the strong/weak side, updating her position as soon as she is executed or she is not placed anymore at the best prices, we define 
$$ 
E^k_\pm := \text{number of trades on the $\pm$ side in }(T_{k-1},T_k] - \indicator{B_k=\pm} ,
$$
where $B_k=J_kJ_{k-1}$ is the same as \refequation{Bn}. Following the strategy $\ell_\pm\equiv 1$, the agent places continuously on one side a limit order of size $L$. We can determine the statistics
$$ 
e^k_\pm(i) := \text{number of agent trades of size $i$ $\in$ $\{1,\ldots,L\}$  on the $\pm$ side in }(T_{k-1},T_k] - \indicator{B_k=\pm,i=L}. 
$$
Then, by setting $E^N_\pm := \sum_{k=1}^{N} E^k_\pm$ and $e^N_\pm\left(i\right) := \sum_{k=1}^{N} e^k_\pm\left(i\right)$,  by the strong law of large numbers,  we have the estimator for 
the probability distribution $\law$:  
$$
 \frac{e_N}{E_N} \longrightarrow   \;  \vartheta_\pm(\{i\},L)   \for\forall i=1,\ldots, L, \;\;\;  \mbox{ as } \; N \rightarrow \infty. 
$$
\ep

\section{Properties of the function $\theta^T(t,s)$} \label{appendix::thetaT}

For the applications of our model to market making problem, we shall extensively use properties of the mean value of the stock price at horizon 
defined in (\ref{pi}).

\subsection{The PDE representation}

We first give a decomposition induced by the arithmetic nature of the stock price and its symmetry.

\begin{Proposition}  \label{lemma::theta}
The function $\pi$  is decomposed into: 
\begin{eqnarray} \label{decpi}
\pi(t,p,i,s)  &=& p +  i \theta(t,s)
\end{eqnarray}
where $\theta$ is  the unique bounded continuous  viscosity  solution to the linear integro-differential equation: 
\begin{equation} \label{appendix::theta}
\begin{cases}
- \Dt{\theta}  - \Ds{\theta}  - \mu(s)   \theta(t,0) +  \sigma^2(s)  \; \theta(t,s) 
- 2 \delta \mu(s) \; = \;  0,  \;\;\; \mbox{ on } \; [0,T)\times\R_+,  \\ 
\theta(T,) \; = \;  0, 
\end{cases}
\end{equation}

\end{Proposition}
\begin{proof} First, we know from standard result on pure jump process (see e.g. \cite{baurie11}) that under continuity of the intensity functions 
$h_\pm$, the function $\pi$ defined by the expectation in \reff{pi} is a continuous function.  Moreover, 
by the Feynman-Kac representation theorem, it is  solution to the linear integro-differential equation:
\begin{equation} \label{linpi}
\begin{cases}
\mpartialts{\pi}  - \sum_{\sym\in\pm} h_\sym(s)\Delta\pi(t,p+2\delta \sym i,\sym i,0) \; = \; 0,  \\
\pi(T,) \;= \; p,
\end{cases}
\end{equation}
for $(t,p,i,s)\in[0,T]\times 2\delta \mathbb Z \times \pmset \times \Rplus$.  Moreover, by considering the functions $\varphi_\pm(t,p)$ $=$ 
$p \pm C(T-t)$, we easily check under the assumption that $h_\pm$ is bounded, that for $C$ large enough, $\varphi_+$ (resp. $\varphi_-$)  is a supersolution (resp. subsolution) to \reff{linpi}. Then,  by  Dynkin's formula, the process $(\varphi_+(t,P_t))_t$ (resp.  $(\varphi_-(t,P_t))_t$) is a 
supermartingale (resp. submartingale), and so: 
\beq
\varphi_-(t,p)  =  p -  C(T-t)  \leq   \E_{t,p,i,s}[\varphi_-(T,P_T)] & = &  \pi(t,p,i,s)  \label{pibound} \\
& = & \E_{t,p,i,s}[\varphi_+(T,P_T)]  \leq   \varphi_+(t,p)  =  p + C(T-t), \nonumber 
\enq
for $(t,p,i,s)\in[0,T]\times 2\delta \mathbb Z \times \pmset \times \Rplus$.  On the other hand, from the additive structure \reff{midprice} of the price process, it is clear that the function $\pi$ defined in \reff{pi} is decomposed into: 
$\pi(t,p,i,s)$ $=$ $p$ $+$ $\kappa(t,i,s)$ for some function $\kappa$ not depending on $p$. 
Let us now set: $\kappa_\pm(t,s)$ $:=$ $\kappa(t,\pm 1,s)$. Then, from \reff{linpi}, we see that $(\kappa_+,\kappa_-)$ is a  bounded pair solution to the system of parabolic  linear integro-differential equations: 
\begin{equation} \label{integrokappa}
\begin{cases}
\mpartialts{\kappa_\pm} + h_-(s)\kappa_\pm  -  h_\pm(s) \Delta \kappa_\pm(t,0)    \mp 2 \delta \mu(s) - h_-(s)\kappa_\mp(t,0)  \; = \; 0, \\
\kappa_\pm(T,.) \; = \; 0. 
\end{cases}
\end{equation}
We observe that $(-\kappa_-,-\kappa_+)$ is also a bounded pair solution to the same system of equations \reff{integrokappa}, and we deduce by uniqueness for such system of 
parabolic  linear integro-differential equations (see again \cite{say91}) that $\kappa_+$ $=$ $-\kappa_-$.  We then set $\theta$ $:=$ $\kappa_+$ so that $\kappa(t,i,s)$ $=$ $i\theta(t,s)$, i.e. \reff{decpi} holds, implying in particular that 
$\theta$ is continuous,  and we clearly see from \reff{integrokappa} that $\theta$ satisfies the equation \reff{appendix::theta}.  Uniqueness of $\theta$ is obtained by observing that if $\tilde\theta$ is another bounded viscosity solution to \reff{appendix::theta}, then $\tilde\pi(t,p,i,s)$ $:=$ $p+i\tilde\theta(t,s)$ is a solution to 
\reff{linpi} with linear growth condition in $p$, and so by uniqueness for such equation: $\tilde\pi$ $=$ $\pi$, hence $\tilde\theta$ $=$ $\theta$. 
\end{proof}

\subsection{The probabilistic representation}

We now investigate further bounds for $\theta(t,s)$, as well as its asymptotic behavior for large horizons, using a probabilistic approach based on the representation (\ref{thetaproba}). 
In order to emphasize the dependence on the horizon $T$, we write $\thetaT(t,s)$ for $\theta(t,s)$ and define the conditional mean of the next jump w.r.t. to the elapsed time as
\begin{eqnarray}
\tilde\alpha(s) &:=& \condprob{E}{}{J_1\condto  I_{0} =+ 1,S_0 = s } \for s\in\Rplus, \labeleq{def_tilde_alpha} \\
&=& \condprob{E}{}{J_1\condto J_{0} =+1,T_1\geq s}.  \label{tildealphaautre}
\end{eqnarray}

\begin{Lemma}\label{lemmaBayes}
For a Markov Renewal Process $(J_k,T_k)$ on a finite state space $E$ with 
\begin{eqnarray*}
Q_{ij} := \condprob{P}{}{J_k=j \vert J_{k-1}=i} \for F_{ij} := \condprob{P}{}{T_k-T_{k-1} \leq s \vert J_k=j, J_{k-1}=i}\for i,j\in E , 
\end{eqnarray*}
we have
\begin{eqnarray}
\condprob{P}{}{J_k = j \vert J_{k-1} = i , T_k-T_{k-1} \geq s} = Q_{ij} \frac{1-F_{ij}(s)}{1- \sum_{j\in E} Q_{ij} F_{ij}(s)} , \;\;\;i,j\in E . \labeleq{_alphatilde}
\end{eqnarray}
\end{Lemma}
\begin{proof}
A simple Bayes argument gives
\begin{eqnarray*}
\condprob{P}{}{J_k = j \vert J_{k-1} = i , T_k-T_{k-1} \geq s} &=&  \frac{\condprob{P}{}{J_k = j , T_k-T_{k-1} \geq s \vert J_{k-1} = i}}{\condprob{P}{}{T_k-T_{k-1} \geq s \vert J_{k-1} = i}}  \\
&=& \frac{\condprob{P}{}{T_k-T_{k-1} \geq s \vert J_{k-1} = i , J_k = j } \condprob{P}{}{J_k=j\vert J_{k-1}=i}}{\sum_{j\in E} \condprob{P}{}{T_k-T_{k-1} \geq s \vert J_{k-1} = i , J_k = j}\condprob{P}{}{J_k=j\vert J_{k-1}=i}} \\
&=& \frac{Q_{ij}\left(1-F_{ij}(s)\right)}{\sum_{j\in E} Q_{ij}\left(1-F_{ij}(s)\right)} \; = \;   Q_{ij} \frac{1-F_{ij}(s)}{1- \sum_{j\in E} Q_{ij} F_{ij}(s)} .
\end{eqnarray*}
\end{proof}

\begin{Proposition}\label{cor::tilde_alpha} 
For $s\in\Rplus$,
\begin{eqnarray} \label{expresstildealpha}
\tilde\alpha(s) &=& \sum_{\sym\in\pm} \sym\left(\frac{1+\sym\alpha}{2}\right)\left(\frac{1-F_{\sym}(s)}{1- F(s)}\right). 
\end{eqnarray}
In particular i) $\tilde{\alpha}(0)=\alpha$, while ii) if marks and tick times are independent, $\tilde\alpha(s)$ $\equiv$ $\alpha$.
\end{Proposition}
\begin{proof}
By  applying Lemma \ref{lemmaBayes} to the Markov renewal process  driving the stock price, hence with  the notation: 
$$
E = \pmset  \for  Q_{ij} = \frac{1+ij\alpha}{2} \for F_{ij} = \sum_{\sym\in\pm} F_\sym 1_{\{ij=\sym\}}, 
$$
we have for any $i$ $\in E$: 
$$
\sum_{j\in E} j \condprob{P}{}{J_1 = j   \vert J_{0} = i , T_1 \geq s}
= \sum_{j\in E} j  Q_{ij} \frac{1-F_{ij}(s)}{1- \sum_{j\in E} Q_{ij} F_{ij}(s)} = \sum_{\sym\in\pm} \sym\left(\frac{1+\sym\alpha}{2}\right)\left(\frac{1-F_{\sym}(s)}{1- F(s)}\right),
$$
which shows \reff{expresstildealpha} from \reff{tildealphaautre}. From \refequation{def_tilde_alpha} and recalling \refequation{Bn}, we see directly that
$\tilde\alpha(0)$ $=$  $\condprob{E}{}{J_1 \vert I_0 = +1,S_0 = 0}$  $=$ $\condprob{E}{}{B_1}$  $=$ $\alpha$. 
Finally,  if  the tick times of  $(N_t)$ and mark process $(J_k)$ are independent, then $F_+ = F_- = F$, and by \reff{expresstildealpha}, we have: 
\beqs
\tilde\alpha(s) &=&  \sum_{\sym\in\pm} \sym\,\frac{1+\sym\alpha}{2} \; = \; \alpha.
\enqs
\end{proof}

 We finally describe the asymptotic behavior of $\theta$ by a standard regenerative argument  for Markov renewal process.

\begin{Proposition}\label{prop::phi_infty}
The asymptotic behaviour of $\theta^T(t,s)$ is given by
\begin{eqnarray*}
\lim_{T\rightarrow\infty} \theta^T(t,s) &=& 2\delta \frac{\tilde\alpha(s)}{1-\alpha} \for \forall s\in\Rplus.
\end{eqnarray*}
\end{Proposition}
\begin{proof}
At time $T_1$ ($T_1<\infty$, since the process $N_t$ is assumed non-explosive), we have: $P_{T_1}$ $=$ $P_0$ $+$ $2\delta J_1$, $I_{T_1} = J_1$,  $S_{T_1} = 0$, and so
\beq \label{interP}
\condprob{E}{}{P_{T_1} \condto P_0=0, I_0=+1,S_0=s} &= & 2\delta \condprob{E}{}{J_{1} \condto I_0=+1,S_0=s } \; = \;  2\delta \tilde\alpha(s). 
\enq
From \reff{midprice} and \refequation{Bn}, we have for all $t$ $\geq$ $T_1$: 
\beqs
P_t  &=&  P_{T_1} + 2\delta \sum_{k=2}^{N_t} {J_k} \; = \;  P_{T_1} + 2 \delta J_1\left( \sum_{k=2}^{N_t}\prod_{i=1}^{k} B_i\right),
\enqs
where $(B_n)_n$ are i.i.d Bernoulli  variables of mean $\alpha$, and independent of $J_1$, which proves that
\beq \label{interP2}
 P_t \condto (T_1,J_1)  \;   \equiv \;  P_{T_1} +  2 \delta J_1 \hat P_{N_t-1}, & \; \mbox{ with } & \hat P_{n}  \equiv   \sum_{k=1}^{n}\prod_{i=1}^{k} B_i, 
\enq
where $(\hat P_{n})_n$ is independent from $J_1$, of mean $\E[\hat P_n]$ $=$ $\sum_{k=1}^n \alpha^k$. 
Now,  from the probabilistic representation \reff{thetaproba} of $\theta$,  and time homogeneity of the price process, we have 
by tower conditioning w.r.t. to $(T_1,J_1)$, and \reff{interP}-\reff{interP2}: 
\beqs
\lim_{T\rightarrow\infty}\theta^T(t,s)  \; = \;  \lim_{t\rightarrow\infty}\theta^t(0,s) &=&  \lim_{t\rightarrow\infty}\condprob{E}{}{P_{t} \condto P_0=0, I_0=+1,S_0=s }  \\
&=&  2\delta\tilde\alpha(s) + 2\delta\tilde\alpha(s) \lim_{n\rightarrow\infty} \mathbb E\big[\hat P_n\big] \\
& = &  2\delta\tilde\alpha(s) + 2\delta\tilde\alpha(s) \sum_{k=1}^{\infty} \alpha^k =  2\delta\tilde\alpha(s) \sum_{k=0}^{\infty} \alpha^k = 2\delta\round{\frac{\tilde \alpha(s)}{1-\alpha}},
\enqs
which ends the proof. 
\end{proof}

\newpage
\bibliographystyle{plain}
\small
\bibliography{total}

\end{document}